\documentclass[apj]{emulateapj}
\usepackage{epsfig}

\newcommand{\beq}{\begin{equation}}
\newcommand{\eeq}{\end{equation}}
\newcommand{\bea}{\begin{eqnarray}}
\newcommand{\eea}{\end{eqnarray}}
\newcommand{\trm}[1]{\textrm{#1}}

\def\cd{\trm{ g cm}^{-2}}
\def\cms{\trm{ cm s}^{-1}}

\def\egs{\trm{ ergs g}^{-1}\trm{ s}^{-1}}

\begin{document}

\shortauthors{WEINBERG \& BILDSTEN}
\shorttitle{CARBON DETONATION IN SUPERBURSTS}

\title{Carbon Detonation and Shock-Triggered Helium Burning\\ in Neutron Star Superbursts}
\author{Nevin N.~Weinberg}
\affil{Astronomy Department and Theoretical Astrophysics Center, 601 Campbell Hall, University of California, Berkeley, CA 94720; nweinberg@astro.berkeley.edu}
\and
\author{Lars Bildsten}
\affil{Kavli Institute for Theoretical Physics and Department of Physics, Kohn Hall, University of California, Santa Barbara, CA 93106; bildsten@kitp.ucsb.edu}

\begin{abstract}
The strong degeneracy of the $^{12}$C ignition layer on an accreting neutron star results in a hydrodynamic thermonuclear runaway, in which the nuclear heating time becomes shorter than the local dynamical time. We model the resulting combustion wave during these superbursts as an upward propagating detonation. We solve the reactive fluid flow and show that the detonation propagates through the deepest layers of fuel and drives a shock wave that steepens as it travels upward into lower density material. The shock is sufficiently strong upon reaching the freshly accreted H/He layer that it triggers unstable $^4$He burning if the superburst occurs during the latter half of the regular Type I bursting cycle; this is likely the origin of the bright Type I precursor bursts observed at the onset of superbursts. The cooling of the outermost shock-heated layers produces a bright, $\approx 0.1\trm{ s}$, flash that precedes the Type I burst by a few seconds; this may be the origin of the spike seen at the burst onset in 4U 1820-30 and 4U 1636-54, the only two bursts observed with {\it RXTE} at high time resolution. The dominant products of the  $^{12}$C detonation are $^{28}$Si, $^{32}$S, and $^{36}$Ar. Gupta et al. showed that a crust composed of such intermediate mass elements has a larger heat flux than one composed of iron-peak elements and helps bring the superburst ignition depth into better agreement with values inferred from observations.   
\end{abstract}

\section{Introduction}
\label{sec:intro} 

Superbursts are powered by unstable thermonuclear burning on the surface of an accreting neutron star in a low mass X-ray binary (for reviews, see \citealt{Kuulkers:04, Cumming:04a, Strohmayer:06}).  Their recurrence time ($\approx 1 \trm{ yr}$) and duration ($\approx \trm{ hours}$) suggest ignition densities of $10^{8}-10^{9} \trm{ g cm}^{-3}$ \citep{Cumming:06}, $100-1000$ times larger than that of normal Type I X-ray bursts. If the superburst fuel is $^{12}$C \citep{Cumming:01, Strohmayer:02}, then the observed energy release ($\approx 10^{42} \trm{ ergs}$) implies a $^{12}$C mass fraction $X_{12} \ga 0.1$ in the matter accumulated from Type I bursts \citep{Cumming:06}. 

Previous studies assumed the entire layer of $^{12}$C fuel burns instantly and hydrostatically, and obtained a good match to the late-time ($\ga10^3\trm{ s}$) light curves and accounted for the quenching of normal bursts in the days following a superburst  \citep{Cumming:01, Strohmayer:02, Cumming:04b, Cooper:05, Cumming:06}. They did not, however, fit the early-time light curves nor provide a mechanism to trigger the Type I bursts that precede superbursts by $\approx10 \trm{ s}$. Such precursor bursts have been found in all five cases in which the onset of the superburst was observed (4U1820-30: \citealt{Strohmayer:02}; 4U 1636-54: \citealt{Strohmayer:02b}; KS 1731-26: \citealt{Kuulkers:02}; 4U 1254-69: \citealt{intZand:03}; GX 17+2: \citealt{intZand:04b}). 

In Weinberg, Bildsten, \& Brown (2006; hereafter WBB), we showed that the strong degeneracy of the superburst ignition layer results in a hydrodynamic thermonuclear runaway, in which the heating time $t_h \equiv (d\ln T_b/ dt)^{-1}$ becomes shorter than the local dynamical time $t_d = h / c_s \approx 10^{-6} \trm{ s}$, where $T_b$, $h$, and $c_s$ are the temperature, pressure scale height, and sound speed at the base of the burning layer.  We start in \S~\ref{sec:hydro} by describing the conditions for the spontaneous initiation of a detonation, and show that the observed superbursts are likely deep enough into the neutron star that their plane-parallel burning will initiate a detonation. We thus model the hydrodynamic combustion wave that forms once $t_h <t_d$ as a detonation, showing its impact on the overlying surface layers. 

We show (\S~\ref{sec:hydro}) that as the detonation propagates upward, it drives an outgoing shock wave that steepens as it travels into lower density material. When the shock impacts the freshly accreted H/He layer, its overpressure is $\Delta p /p \approx 10$. Our calculations (\S~\ref{sec:trigger}) suggest that about half the time there is enough H/He present in the layer that the shock can ignite the $^4$He and trigger a Type I precursor burst. We compute full light curves in \S~\ref{sec:lightcurves} and show that a detonation with a shock-triggered Type I burst can explain features of the early-time light curves while still accounting for the late-time behavior. We also show that the cooling of the outermost shock-heated layers results in a bright, sub-second, flash that precedes the precursor by a few seconds. This could be the origin of the spike seen at the burst onset in 4U1820-30 and 4U 1636-54, the only two bursts observed with {\sl RXTE} at high time resolution. 

We show in \S~\ref{sec:lightcurves} that even if the shock does not ignite the $^4$He, the cooling of the ashes of $^{12}$C burning at depths just below the H/He layer can result in a precursor that appears similar to a Type I burst, but with a peak luminosity that is smaller by a factor of $\simeq 2-3$. This may explain why three out of the five precursor bursts were only about half as bright as the ordinary Type I bursts \citep{Kuulkers:04, Cumming:06}.

\section{Hydrodynamic Carbon Burning} 
\label{sec:hydro}

After the thermally unstable ignition of $^{12}$C, there is an hour-long convective stage where the energy generation rate from $^{12}$C burning, $\epsilon_{\rm C}$, is sufficiently small that the heating time $t_h$ is much longer than the eddy turnover time $t_e = h / v_c$, where $v_c=v_c(T_b)$ is the typical velocity of a convective cell (WBB). For low $^{12}$C ignition depths (i.e., low densities) or initial $^{12}$C mass fractions $X_{12}$, the $^{12}$C completely burns in this convective stage, and hydrostatics describes all stages of burning (just as with $^4$He burning during Type I X-ray bursts; \citealt{Weinberg:06}).

\begin{figure}
\epsfig{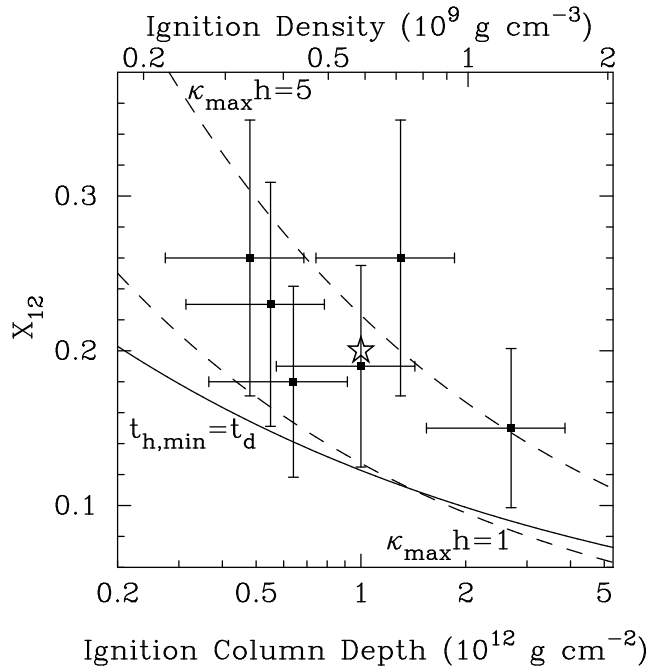}
\caption{Range of superburst ignition parameters $y_b$ and $X_{12}$, with the top axis showing the base density at ignition, $\rho_b$. The region above the solid line has a minimum heating time $t_{\rm h, min}$ less than the dynamical time $t_d$. The six data points are from fits by \citet{Cumming:06} to observed superburst light curves. The error bars assume a 30\% uncertainty in distance to each source (and use the scalings with distance given in \citealt{Cumming:06}).  The open star at $y_b=10^{12}\cd$, $X_{12}=0.2$, indicates the parameters chosen for the detonation calculation shown in Figures 2--5. The upper (lower) dashed curve shows where $\kappa_{\rm max} h =5$ ($\kappa_{\rm max} h =1)$. \label{fig:X12yb}}
\end{figure}

However, for larger $^{12}$C ignition depths $y_b\ga2\times10^{11}\cd$ (ignition densities\footnote{Assuming a relativistic degenerate electron gas, the density at the base is $\rho_b \approx (5\times10^8\trm{ g cm}^{-3})(g_{14}/2)^{3/4}(0.5/Y_e)y_{b,12}^{3/4}$ where $g_{14}=g/10^{14} \trm{ cm s}^{-2}$ is the gravity, $Y_e$ is the electron fraction, and $y_{b,12}=y_b/10^{12}\cd$.} $\rho_b\ga1.5 \times 10^8 \trm{ g  cm}^{-3}$) and $X_{12}\ga0.2$, $\epsilon_{\rm C}$ becomes so large that $t_h < t_e$ and convection becomes inefficient at transporting the energy release outward ($v_c\simeq0.07c_s$ when $t_h=t_e$ for $y_b=10^{12}\cd$, $X_{12}=0.2$; WBB). As a result, the fuel is rapidly consumed within a thin layer near the base in a local thermonuclear runaway, during which time $t_h$ becomes shorter than the dynamical time $t_d$ and hydrodynamic burning begins. The transition from $t_h=t_e$ to $t_h=t_d$ only requires a $\approx 20$\% increase in temperature (WBB).  We calculate the minimum heating time $t_{\rm h, min}$ using the estimate in WBB (equation [5]) and in Figure \ref{fig:X12yb} plot the relation between $y_b$ and $X_{12}$ where $t_{\rm h, min}=t_d$. Ignitions above this line will become hydrodynamic.  

The only direct information we have about $y_b$ and $X_{12}$ is that derived by \citet{Cumming:06} from fits to six observed superburst light curves. These are plotted in Figure \ref{fig:X12yb} assuming a 30\% uncertainty in the distance to each source (corresponding to the uncertainty in the distance to 4U 1254-690, the only one of the six sources with a reported best fit distance with error bars).\footnote{The light curves constrain the nuclear energy release $E_{17}=E_{\rm nuc}/10^{17}\trm{ ergs g}^{-1}$. To estimate $X_{12}$ we assume $X_{12}=0.1 E_{17}$ corresponding to $^{12}$C burning to $^{56}$Fe.  As we show in \S~\ref{sec:upward}, however, the $^{12}$C may only burn to $^{28}$Si, in which case $X_{12}$ is larger by $\approx 40\%$.} The measured values of $y_b$ and $X_{12}$ suggest that the $^{12}$C likely burns hydrodynamically. We argue in \S~\ref{sec:spontaneous} that a possible (but by no means proven) outcome is a spontaneous initiation that leads to a propagating detonation.

\subsection{Spontaneous Initiation of a Detonation}
\label{sec:spontaneous}

Even for chemical combustion in gases, it remains challenging to theoretically predict the conditions needed for a spontaneous detonation \citep{Lee:77, Lee:84, Bdzil:07}. However, a widely accepted criteria for initially addressing the question is the requirement that a sufficiently large volume (defined below) burn in less than a sound crossing time \citep{Blinnikov:86,Blinnikov:87,Woosley:90,Niemeyer:97, Khokhlov:97}. 

We first estimate the size of the region that burns in less than a sound crossing time. According to the Zel'dovich criterion (see \citealt{Bartenev:00} for an updated discussion), such burning requires a spatial region where the temperature gradient is so shallow that nearly simultaneous runaways can occur. The quantitative criterion is that the phase speed of the burning $v_{\rm ph}=|dt_{\rm nuc} / dx|^{-1}>c_s$, where $t_{\rm nuc} = (d\ln \epsilon_C/dt)^{-1}=t_h/\nu \approx C_p T/\nu\epsilon_C$ is the time scale for the acceleration of nuclear burning and $\nu=d\ln\epsilon_C/d\ln T\approx 21$ at $T=2\times 10^9 {\rm K}$. We first evaluate this in the vertical direction assuming an adiabatic temperature gradient at the moment when $t_{\rm nuc}=t_e$, which occurs at $T_b\approx2\times10^9\trm{ K}$ (we integrate over the entire convective region when computing $t_{\rm nuc}$; see Figure 1 and equation [3] in WBB). Since the heat capacity is determined mainly by the electrons, $C_p \propto T/y^{1/4}$ \citep{Cumming:01}, and thus $t_{\rm nuc}\propto T^{2-\nu}/y$. Requiring that $v_{\rm ph}>c_s$ then implies
\beq 
t_h(y) <{\lambda}t_d\approx 3t_d, 
\eeq
where $\lambda=\nu/[1+(\nu-2)\nabla_{\rm ad}]$ and $\nabla_{\rm ad}=(d\ln T/d\ln y)_{\rm ad}\approx0.3$ is the adiabatic index. At runaway $t_h(y_b)=t_d$, and $v_{\rm ph}>c_s$ between the base and a height 
\beq 
\label{eq:zcrit}
z_{\rm crit} \approx h\ln(y_b/y_{\rm crit}) 
\approx h\frac{\lambda\ln\lambda}{\nu}\approx 0.2h.  
\eeq
Above that location, the temperature is not adequate to help initiate the detonation. Of course, because of the small scale height compared to the radius, there is a much smaller temperature gradient in the horizontal direction, an issue we will address momentarily.

If the initiating region calculated above is too small, the initial shock is unable to trigger further burning before it decays by geometrical dilution. Direct numerical simulations and experiments always find that to trigger a detonation the initiating region must be a few orders of magnitude longer than the induction length of the detonation $l_{\rm ind}\approx v_{CJ} t_{\rm ind}$ (see \citealt{Eckett:00} for a recent discussion). Here $v_{CJ}$ is the Chapman-Jouguet velocity (see e.g., \citealt{Khokhlov:89}) and $t_{\rm ind}$ is the induction time, the time scale to complete the burning behind the shock. In the case of a spherically symmetric ignition (see \citealt{He:94}), such as in a centrally ignited Type Ia supernova, this minimum distance is expressed in terms of a maximum radius of curvature $\kappa_{\rm max}$ \citep{Sharpe:01, Dursi:06}. \citet{Dursi:06} carried out direct numerical simulations for carbon detonations, and derived a fitting function for $\kappa_{\rm max}(\rho, X_{12})$.  The most conservative estimate of the critical conditions needed for a detonation is to then demand that the \emph{vertical} length of the spontaneously igniting region, $z_{\rm crit}$, be larger than $\kappa_{\rm max}^{-1}$. By equation (\ref{eq:zcrit}), this translates into a requirement that $\kappa_{\rm max} h >5$, as plotted in Figure \ref{fig:X12yb}. Most of the data lie below this line, making it challenging to trigger a spherical detonation by a vertical initiation.

Building on the arguments for Type Ia supernovae core detonations, our discussion has assumed that the vertical temperature gradient (where the temperature decreases by $\approx 30-40\%$ over a pressure scale height $h$) plays the critical role. However, the superburst occurs in a thin shell at the surface of the neutron star, and the hour long convective phase prior to the hydrodynamic runaway should be adequate to establish transverse temperature gradients smaller than the vertical gradient. Only a numerical anelastic calculation could accurately calculate such a gradient, but we think it reasonable to assume that the initiation length is larger in the plane, potentially exceeding $h$ at some location. Such a cylindrical geometry also has the advantage of less severe geometrical dilution, where all analytic calculations yield a factor of two smaller value for $\kappa_{\rm max}^{-1}$ \citep{Lee:77, He:94}. For this reason, we plot an additional line at $\kappa_{\rm max} h=1$ in Figure \ref{fig:X12yb}, representing our current best guess of the constraint for detonation initiation in a planar geometry. All the data points are above this line, as is the explosion we are simulating in this paper (shown by the open star). Although more work clearly remains to fully assess the onset of a detonation in a planar geometry after a long convective phase, we will proceed assuming that it can happen.

\subsection{Numerical Method}
\label{sec:numerical}
We assume a plane parallel geometry with constant gravity $g=2.4\times10^{14}\trm{ cm s}^{-2}$ (corresponding to $M=1.4M_\odot$, $R=10\trm{ km}$) and treat the detonation in only the vertical direction, parametrized by the column depth $y$ into the star (units of $\trm{g cm}^{-2}$), where $dy = -\rho dr$. Modeling the lateral propagation is important and left for future study (see \citealt{Zingale:01}, who consider a laterally propagating $^4$He detonation in the ocean of a neutron star). We solve the reactive fluid flow equations using a one-dimensional, explicit, Lagrangian, finite difference scheme (see e.g., \citealt{Benz:91}). We use operator splitting to couple the hydrodynamics to a nuclear energy generation network. The network contains 13 isotopes: $\{^4$He, $^{12}$C, $^{16}$O, $^{20}$Ne, $^{24}$Mg, $^{28}$Si, $^{32}$S, $^{36}$Ar, $^{40}$Ca, $^{44}$Ti, $^{48}$Cr, $^{52}$Fe, $^{56}$Ni$\}$, and includes $\alpha$-chain, heavy-ion, and $(\alpha, p)(p, \gamma)$ reactions\footnote{We use F. Timmes' network solver ``public\_aprox13.f", kindly made available online at http://www.cococubed.com/code\_pages/net\_aprox13.shtml (see also \citealt{Timmes:99}).}. We assume electrons, ions, and photons supply the pressure and calculate the equation of state, volumetric neutrino emissivity, and thermal conductivity, as in \citet{Brown:04}. 

We model the detonation and shocks over the column depths $10^4-10^{14}\trm{ g cm}^{-2}$ using 4000 grid points spaced uniformly in $\log y$. The $^{12}$C layer spans the range $y_{\rm He} < y < y_b \sim 10^{12} \cd$. The depth of the base of the freshly accreted H/He layer, $y_{\rm He}(\phi) \equiv \phi y_{\rm He}(\phi = 1)  \sim10^8\cd$, varies linearly with the phase $\phi=t/t_{\rm recur}$ of the Type I burst cycle, whose recurrence time $t_{\rm recur} \sim 10^4-10^5 \trm{ s}$. We assume the accreted gas has a solar composition and consider local accretion rates per unit area in the range $\dot{m}=0.05-0.3 \dot{m}_{\rm Edd}$, where $\dot{m}_{\rm Edd}$ is the local Eddington accretion rate (i.e., roughly the range inferred from observations; \citealt{Kuulkers:04}). The $^{12}$C layer has a mass fraction $1-\trm{X}_{12}$ of $^{56}$Fe and for $y>y_b$ the composition is pure $^{56}$Fe. For a given $y_b$, $X_{12}$, $\phi$, and $\dot{m}$, we solve for the crustal heat flux that results in unstable ignition at $y_b$ (see WBB) and then calculate $y_{\rm He}(\phi=1, \dot{m})$ as in \citet{Cumming:00}. 

The shock reaches the top of our grid ($y=10^4\cd$) at a time $t_{\rm top}\approx10\trm{ $\mu$s}$ after the onset of the hydrodynamic runaway. The underlying layers are adiabatically expanding at $t=t_{\rm top}$, though they remain bound to the neutron star and in plane parallel (see \S~\ref{sec:steepening}). We assume that for $t>t_{\rm top}$ the flow remains adiabatic and that after a few dynamical times (tens of $\mu\trm{s}$), the shocked layers settle into a new, puffed-up, hydrostatic equilibrium, with an entropy profile given by that at $t=t_{\rm top}$. This post-shock hydrostatic equilibrium sets the early-time light curve of our cooling calculations (\S~\ref{sec:lightcurves}). 

\subsection{Upward propagating detonation}
\label{sec:upward}

\begin{figure}
\begin{center}
\epsfig{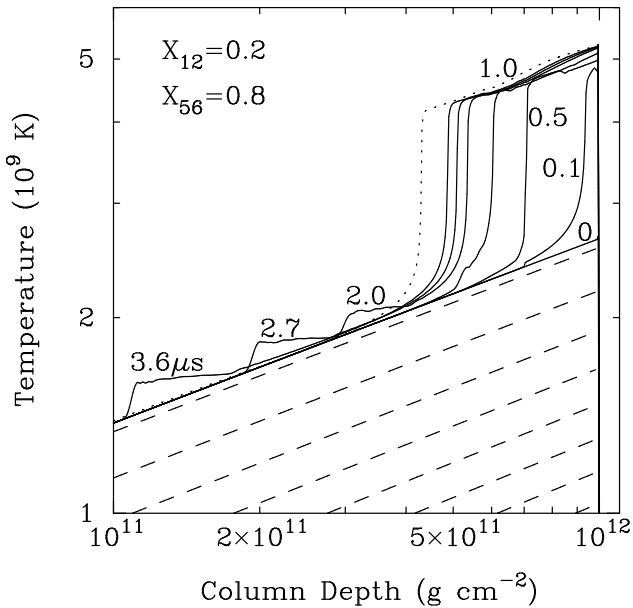}
\end{center}
\caption{Evolution of the thermal profile for $y_b = 10^{12} \cd$ and an initial composition of 20\% $^{12}$C and 80\% $^{56}$Fe by mass. The sequence of dashed lines show the evolution during the convective stage as in WBB. The sequence of solid curves (with times labeled in microseconds) show the outward propagating detonation which is gradually outrun by the steepening shock. The dotted line shows the profile after $9\trm{ $\mu$s}$, indicating the maximum extent of the detonation $y_{\rm det}$ ($=4.3\times10^{11}\cd$ in this case). \label{fig:Tyzoom}}
\end{figure}

We compute the evolution during the convective stage ($t_h > t_d$) as in WBB. Approximately 30\% of the $^{12}$C is burned to $^{24}$Mg during the convective stage for $y_b=10^{12}\cd$ and a pre-ignition mass fraction $X_{12}=0.2$. We then initiate the detonation at the onset of the dynamical runaway ($t_h = t_d$) by increasing the temperature of the grid point at $y_b$ by 1\%. In Figure \ref{fig:Tyzoom} we show the evolution of the thermal profile during the convective stage as $T_b$ rises ({\sl dashed lines}) and the hydrodynamic stage as the detonation propagates outward ({\sl solid lines}).

The shock wave that defines the head of the detonation front is sufficiently strong that it triggers complete $^{12}$C burning in the deepest layers. However, as the shock propagates outward into cooler, lower density fuel, it fails to trigger $^{12}$C burning on the hydrodynamic timescale ($\sim\mu\trm{s}$). The shock ultimately races ahead of the burning and the detonation dies at a depth  $y_{\rm det}$ (see {\sl dotted line} in Figure \ref{fig:Tyzoom}). The value of $y_{\rm det}$ is approximately the depth at which the detonation induction length $l_{\rm ind}$ becomes greater than a scale height $h = y/\rho$ (this assumes the detonation is very nearly planar upon reaching this depth; see discussion in \S~\ref{sec:spontaneous}).

\begin{figure}
\begin{center}
\epsfig{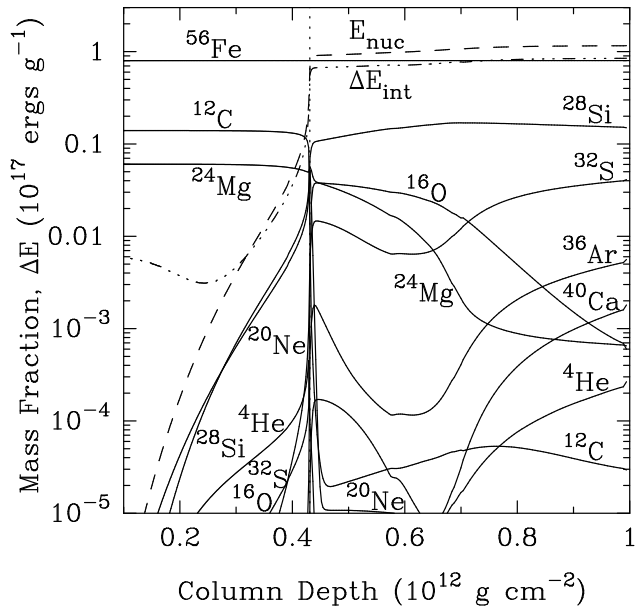}
\end{center}
\caption{Composition profile (\emph{solid lines}), nuclear energy release $E_{\rm nuc}$ (\emph{dashed line}), and increase in internal energy $\Delta E_{\rm int}$ (\emph{dash-triple-dot line}), after traversal of the shock in the detonation region $y_{\rm det} < y < y_b$ for $y_b = 10^{12} \cd$ and an initial composition $X_{12}=0.2$, $X_{56}=0.8$. At depths $y < y_{\rm det}= 4.3 \times 10^{11} \cd$ the detonation fails to trigger $^{12}$C burning   and as the shock steepens $\Delta E_{\rm int}$ becomes greater than $E_{\rm nuc}$. The $^{12}$C at $y<y_{\rm det}$ will later burn in a deflagration (\S~\ref{sec:lightcurves}).\label{fig:yXzoom}}
\end{figure}

We now estimate how $y_{\rm det}$ depends on the ignition parameters. The detonation Mach number $M=v/c_s\approx1.3$ and is close to the Chapman-Jouguet value $M_{\rm CJ}=(1+\alpha)^{1/2}+\alpha^{1/2}$, where $\alpha=(\gamma^2-1)E_{\rm nuc}/2 \gamma h g\simeq0.1$, $\gamma=4/3$, and $E_{\rm nuc}=10^{17} \trm{ ergs g}^{-1}$ \citep{Landau:59}. Since $t_{\rm ind}\approx t_{\rm h, min}$, the minimum nuclear heating time (WBB), we have $l_{\rm ind} = v t_{\rm ind} \approx c_s t_{\rm h, min}$, so that $l_{\rm ind}=h$ when $t_{\rm h, min} = t_d$. In WBB we derived an estimate of the depth $y_{\rm dyn}$ where $t_{\rm h, min}$ becomes greater than $t_d$; thus $y_{\rm det}\approx y_{\rm dyn}$, which gives
\beq
y_{\rm det} \approx (2.4\times10^{11}\cd) \left(\frac{0.2}{X_{12}}\right)^{3.2} \left(\frac{Y_e}{0.5}\right)^{3.0} 
\left(\frac{2}{g_{14}}\right)^{0.7}.
\eeq
 
In Figure \ref{fig:yXzoom} we show the composition profile after traversal of the shock in the detonation region $y_{\rm det} < y < y_b$ for $y_{b,12}=y_b/10^{12}\cd =1$. We find that for $y_{b,12}=0.5$ and $y_{b,12}=1$, the dominant products of burning are the intermediate mass elements $^{28}$Si and $^{32}$S. For $y_{b,12}=5$, the dominant products are $^{28}$Si and $^{32}$S in the top half of the detonation layer's mass and $^{56}$Ni in the bottom half. As we discuss in \S~\ref{sec:summary}, the increased crustal heat flux from a crust composed of $^{28}$Si and $^{32}$S can help bring the superburst ignition depth into better agreement with values inferred from observations \citep{Gupta:06}. 

We also show in Figure \ref{fig:yXzoom} the magnitude of the nuclear energy release in the detonation $E_{\rm nuc} \simeq 1.1\times10^{17} \trm{ ergs g}^{-1}$ and the change in internal energy of the material within the detonation region $\Delta E_{\rm int} \simeq 0.8\times10^{17}\trm{ ergs g}^{-1}$. Here $\Delta E_{\rm int}=E_{{\rm int}, f}-E_{{\rm int}, 0}$, the difference between the internal energy,  $E_{{\rm int}, f}$, upon hydrostatic settling after the detonation passes, and the internal energy,  $E_{{\rm int}, 0}$, at runaway. Thus, a fraction $1-\left(\Delta E_{\rm int} / E_{\rm nuc}\right) \simeq 0.3$ of the nuclear energy release is used to power the upward and downward propagating shock. The downward propagating shock is weak ($\Delta p / p < 0.1$) and will penetrate to great depths before dissipating. However, the total energy deposited $< 0.03 \trm{ MeV nucleon}^{-1} \ll Q_{\rm crust}$, the energy per nucleon released in the crust from pycnonuclear and electron capture reactions. Thus, the shock is not a significant source of crustal heating.
 
The boundary at $y_{\rm det}$ between the hot ashes and the overlying unburned $^{12}$C fuel is convectively unstable. All the $^{12}$C between $y_{\rm He} < y < y_{\rm det}$ will ultimately burn in a convective deflagration, perhaps to iron-peak elements (\S~\ref{sec:lightcurves}). Nonetheless, the contrast persists for more than a dynamical time and thus does not affect the upward propagating shock. 

\subsection{Steepening of upward propagating shock}
\label{sec:steepening}
\begin{figure}
\begin{center}
\epsfig{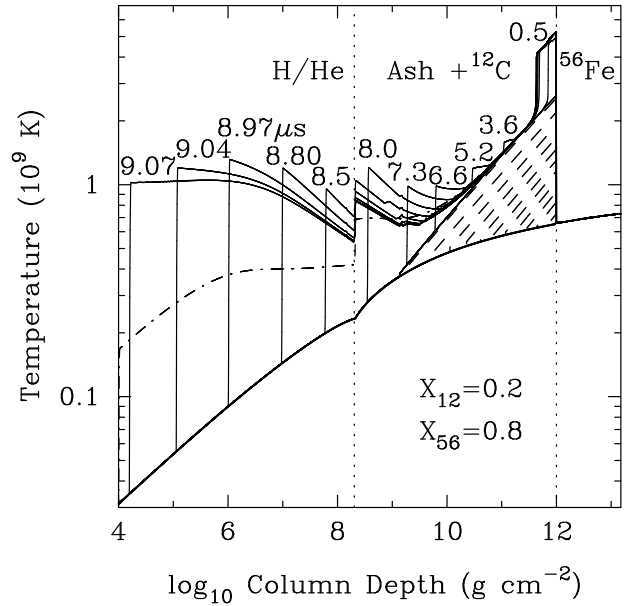}
\end{center}
\caption{Evolution of the thermal profile as the shock continues to propagate outward (sequence of solid lines, labeled in $\mu$s) for the same run as Figure \ref{fig:Tyzoom}. The sequence of dashed lines show the evolution during the convective stage. The vertical dotted lines show the locations of the H/He---ash/C and ash/C---Fe interfaces. The dashed-dotted line shows the profile after the shocked layers have adiabatically expanded and settled into a new hydrostatic solution. The temperature jump due to the shock propagating  downward into the $^{56}$Fe is too small to be seen on this scale.\label{fig:Ty}}
\end{figure}

\begin{figure}
\begin{center}
\epsfig{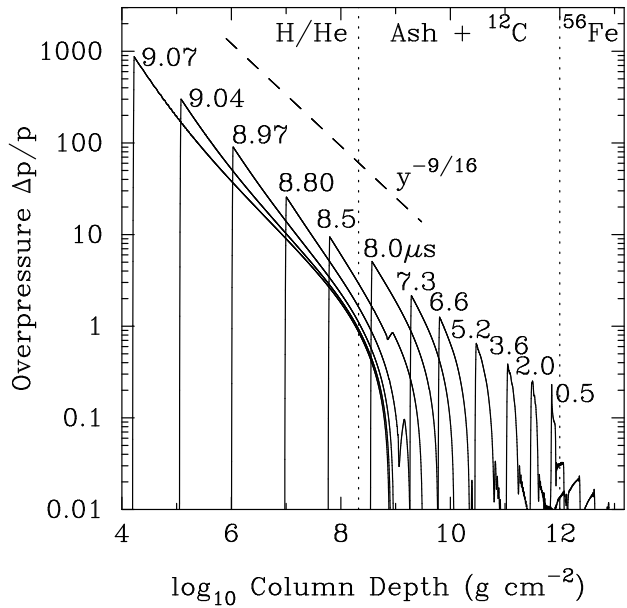}
\end{center}
\caption{Evolution of the shock overpressure for the same run as the previous figures (we only plot $\Delta p /p >0$ and thus do not show the rarefaction wave that trails the shock). The shock steepens as roughly $y^{-9/16}$. The vertical dotted lines show the locations of the H/He---ash/C and ash/C---Fe interfaces. \label{fig:dpp}}
\end{figure}

In Figures \ref{fig:Ty} and \ref{fig:dpp} we show the evolution of the temperature profile and shock overpressure $\Delta p / p = (p_{\rm sh} - p_0) / p_0$ as the shock propagates upward (and downward). WBB showed that as the shock travels down the density gradient it steepens as $\Delta p / p \propto y^{-9/16}$. Within the detonation region ($y_{\rm det} < y < y_{\rm b}$) we have $\Delta p / p \approx E_{\rm nuc} / E_{\rm int, 0}\approx 0.1$. This energy release powers an upward and downward propagating shock such that for $y<y_{\rm det}$ and $y>y_b$,
\bea
\label{eq:dpp}
\frac{\Delta p}{p} 
&\approx& \frac{1}{2}\left(\frac{E_{\rm nuc}}{E_{\rm int, 0}}\right)\left(\frac{y}{y_b}\right)^{-9/16} \nonumber\\ 
&\approx&9 \left(\frac{0.5}{Y_e}\right)\left(\frac{2}{g_{14}}\right)^{1/4}
\left(\frac{y_b}{10^{12}\cd}\right)^{5/16}\nonumber \\ 
& &\times\left(\frac{X_{12}}{0.2}\right) 
\left(\frac{y}{10^8\cd}\right)^{-9/16},
\eea
where the factor of $1/2$ in the first expression accounts for the two shocks.  The numerical estimate assumes a relativistic degenerate electron gas within the detonation region with $E_{\rm int,0}=3 Y_e E_{\rm F}/4=(10^{18} \trm{ ergs g}^{-1}) (g_{14}y_{12})^{1/4}(Y_e/0.5)$, where $E_{\rm F}$ is the Fermi energy, and $E_{\rm nuc}=(6\times10^{17} \trm{ ergs g}^{-1})X_{12}$.  Thus, $\Delta p/p \approx 10$ upon reaching the H/He layer at $y_{\rm He}\approx 10^8 \cd$. 

When the shock impacts the density discontinuity at the H/He---ash/C interface $y_{\rm He}$, a rarefaction wave forms and reflects back downward into the previously shocked material (see e.g., the curve labeled $8.5\trm{ $\mu$s}$ in Figure \ref{fig:dpp}). The upward propagating shock weakens slightly ($\Delta p / p$ decreases by $\approx 20-30\%$) and a steep temperature gradient forms at $y_{\rm He}$. However, despite the steepness of the gradient, the fluid is not convectively unstable because the interface's initial density contrast is maintained in the shock's wake. 

The dashed-dotted line in Figure \ref{fig:Ty} shows the temperature profile after passage of the shock and hydrostatic readjustment. Since $\Delta p / p$ is small for $y \ga 10^{10}\cd$, these layers barely need to expand and cool in order to regain hydrostatic equilibrium. By contrast, the shock is strong for $y \la 10^{8}\cd$ and these layers expand and cool significantly. In \S~\ref{sec:lightcurves} we show that the cooling of the region $10^6\cd \la y \la 10^{8}\cd$ results in a brief ($\approx 0.1\trm{ s}$), bright, flash that precedes the precursor burst by several seconds. The duration of the flash is determined by the hydrostatic temperature profile in this region, which we find is nearly isothermal for the following reason. The layer is initially radiative $T_0 \propto p_0^{1/4}\propto y^{1/4}$, and the shock is gas pressure-dominated, $T_{\rm sh}/T_0 \propto p_{\rm sh} / p_0$. Thus, equation (\ref{eq:dpp}) gives $T_{\rm sh} \propto y^{-5/16}$. The layer then adiabatically expands from $p_{\rm sh}$ back to $p = p_0=gy$ (the layer always remains plane-parallel) and the temperature decreases to a hydrostatic value that varies weakly with depth $T_{\rm hse} \approx T_{\rm sh} (p_0 / p_{\rm sh})^{2/5} \propto y^{-7/80}$.

Near the top of our grid, the shock becomes radiation pressure dominated with a post-shock temperature $T_{\rm sh} \approx (3 \Delta p /a)^{1/4} \propto y^{7/64}$. The diffusion time of the photons generated in the shock $t_{\rm diff} \sim \kappa y h / c$ is larger than the shock propagation time $t_{\rm sh} \sim h / v_{\rm sh}$ as long as $y \ga c / \kappa v_{\rm sh}$, where the opacity $\kappa\sim 0.2 \trm{ cm}^2 \trm{ g}^{-1}$. Thus, photons do not preheat the gas ahead of the shock until $y\la10^2 \cd$ and radiative transport can be neglected over the range of our shock calculation.

For a strong shock, $M^2=(v_{\rm sh}/c_s)^2 \approx \Delta p / p$. In the radiative outer layers $c_s \approx (10^8\cms) y_6^{1/8}$ so that $v_{\rm sh} \approx (10^9\cms) y_6^{-5/32}$, where $y_6 = y/10^6 \cd$ and we used (\ref{eq:dpp}). Since the neutron star escape speed $v_{\rm esc} \approx c/3$, the shock ejects only the outermost layers $y \la 1 \cd$. 

\section{Hydrogen and Helium Burning} 
\label{sec:trigger}

As the shock propagates through the upper layers, it deposits considerable entropy. In \S~\ref{sec:unstableHeburn} we show that if the superburst occurs during the latter half of the Type I bursting cycle, there is enough $^4$He present in the H/He layer that the shock triggers \emph{unstable} $^4$He burning. The $^4$He is consumed within only a few seconds and, as we show in \S~\ref{sec:precursor}, this rapid energy release results in a nearly Eddington-limited Type I burst. In \S~\ref{sec:stableHHeburn} we show that if the superburst occurs during the first half of the cycle, the $^4$He still burns, but only in a thermally stable manner, over a duration of $\ga100\trm{ s}$. Such stable $^4$He burning does not result in a Type I burst because the $^4$He is consumed so slowly that the heat flux from burning never gets very high (though we show that stable rp-process H burning may increase the flux considerably). 
 
\subsection{Shock-Triggered Unstable Helium Burning}
\label{sec:unstableHeburn}
The shock will trigger unstable $^4$He burning only if the base of the $^4$He layer is thermally unstable \emph{after} settling into its post-shock hydrostatic configuration. This is because the shocked layers hydrostatically settle within only a few dynamical times ($\sim 10 \mu\trm{s}$). By contrast, the heating timescale from unstable $^4$He burning at these depths is always greater than $\sim 1 \trm{ ms}$ (see e.g., \citealt{Weinberg:06}). 

\begin{figure}
\begin{center}
\epsfig{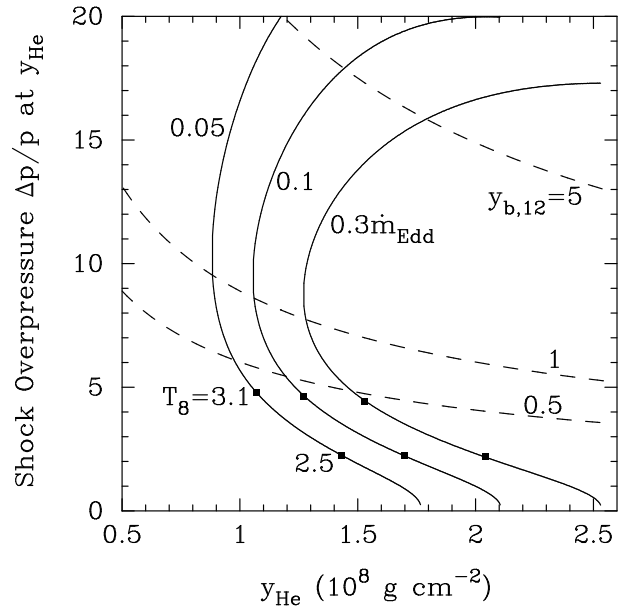}
\end{center}
\caption{The $^4$He ignition curve (\emph{solid lines}) as a function of the shock overpressure $\Delta p/p$ at $y_{\rm He}$ for $\dot{m}/\dot{m}_{\rm Edd}=0.05, 0.1,$ and 0.3. The dashed lines indicate the overpressure at $y_{\rm He}$ for a $^{12}$C detonation ignited at $y_{b,12}=0.5, 1,$ and 5 with $X_{12}=0.2$. The two squares along each curve indicate a Type I burst cycle phase of $\phi=0.6$ and $\phi=0.8$. The squares on the $0.05\dot{m}_{\rm Edd}$ curve are labeled by the temperature $T/10^8\trm{K}$ at $y_{\rm He}$ after hydrostatic settling. \label{fig:trigger}}
\end{figure}

In Figure \ref{fig:trigger} we show the hydrostatic post-shock $^4$He ignition curve as a function of $\Delta p / p$ at $y_{\rm He}$. We define ignition according to the one-zone thermal instability criterion $d\epsilon_{3\alpha}/dT = d\epsilon_{\rm cool}/dT$ (\citealt{Fujimoto:81, Fushiki:87, Cumming:00}), where $\epsilon_{3\alpha}$ is the triple-alpha energy generation rate (we use that from \citealt{Fushiki:87}) and $\epsilon_{\rm cool}=\rho K T/y^2$ is an approximation to the local cooling rate for a thermal conductivity $K$. To the right of the ignition curve $d\epsilon_{3\alpha}/dT > d\epsilon_{\rm cool}/dT$ and burning is unstable. The three curves are for different local accretion rates per unit area $\dot{m}$, in units of the local Eddington rate $\dot{m}_{\rm Edd}$. The lower the $\dot{m}$, the larger the $^4$He abundance at $y_{\rm He}$ \citep{Cumming:00}, and hence the lower the $\Delta p/p$ threshold for ignition. If the last Type I burst was too recent (i.e., the phase, and thus $y_{\rm He}$, is too small), there is no value of $\Delta p / p$ capable of triggering unstable burning since at low densities cooling is always faster than heating \citep{Fujimoto:81}.  If $\Delta p / p$ is too large, $^4$He burning is always stable and will not produce a Type I burst; however, such a strong shock is unlikely since it requires $y_b>5\times10^{12}\cd$ and a superburst early in the Type I burst cycle. 

For a $^{12}$C detonation at $y_b >5\times10^{11} \cd$, the shock is sufficiently strong to trigger unstable $^4$He burning for the latter $\approx40\%- 50$\% of the burst cycle. If a less massive layer of $^4$He is present, a Type I burst will not be triggered. 

A precursor has been seen in each of the five cases where the onset of the burst was detected, and it is unlikely that all are due to shock-triggered $^4$He burning. However, as we describe in \S~\ref{sec:lightcurves}, some of the observed precursors may not be due to unstable $^4$He burning but rather  to $^{12}$C burning at depths just below $y_{\rm He}$. Furthermore, we may be underestimating the fraction of superbursts that trigger unstable $^4$He burning for two reasons. First, the H/He layer may be more He-rich than our calculations assume, especially at higher $\dot{m}$. This is because superbursts tend to occur in systems with large values of $\alpha$ ($\ga 1000$), the ratio of the time-averaged persistent to burst luminosity \citep{intZand:04}. The Type I bursts in these systems also tend to be of short-duration, consistent with a pure $^4$He flash.  On the other hand, the large $\alpha$ values may also indicate that most of the $^4$He is burning stably \citep{Narayan:03, Cooper:05, Cooper:06}, and it is not clear how a shock would trigger a transition from stable to unstable burning. Second, if a convective $^{12}$C deflagration propagates outward from $y_{\rm det}$ and reaches $y_{\rm He}$, it may also trigger unstable $^4$He burning. Such a deflagration would propagate at roughly the turbulent speed $v_{\rm turb} \ga 10^6 \cms$ (cf. \citealt{Timmes:00b}) and reach the $^4$He layer a few milliseconds after the shock. 

\subsection{Stable Hydrogen and Helium Burning}
\label{sec:stableHHeburn}

\begin{figure}
\begin{center}
\epsfig{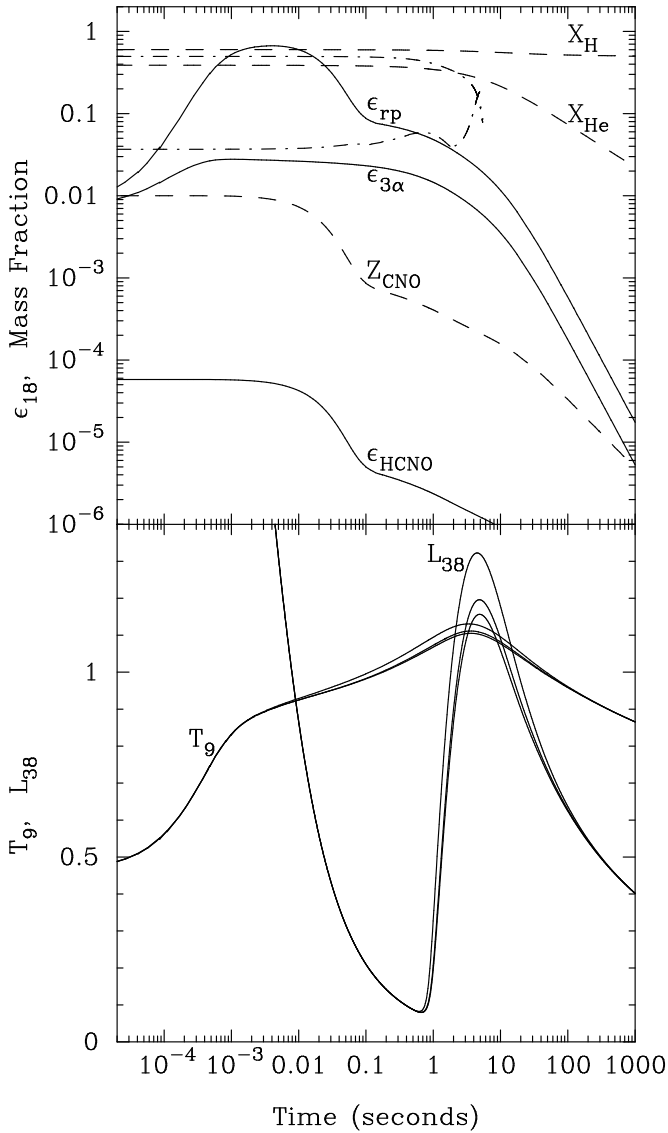}
\end{center}
\caption{Stable H and $^4$He burning at the base of the H/He layer $y_{\rm He}$ as a function of time after hydrostatic readjustment for $y_{b,12}=1$,  $\dot{m}=0.1\dot{m}_{\rm Edd}$, and a Type I burst phase $\phi = 0.5$ (i.e., the shock does not trigger unstable $^4$He burning). The top panel shows the energy generation rate ({\it solid lines}; in units of $10^{18} \trm{ ergs g}^{-1}\trm{ s}^{-1}$) due to rp-process burning $\epsilon_{\rm rp}$ (equation [\ref{eq:rpburning}]), stable triple-alpha $^4$He burning $\epsilon_{3\alpha}$, and hot CNO burning $\epsilon_{\rm HCNO}$. The dashed lines show the mass fractions $X_{\rm H}$, $X_{\rm He}$, and $Z_{\rm CNO}$. Shown for comparison ({\it dash-dot lines}) are $\epsilon_{3\alpha}$ and $X_{\rm He}$ during the convective stage of unstable $^4$He burning for the case $y_{b,12}=1$,  $\dot{m}=0.1\dot{m}_{\rm Edd}$, and $\phi=1$. The bottom panel shows the temperature $T_9 = T / 10^9\trm{ K}$ at  $y_{\rm He}$, and the luminosity $L_{38}=L / 10^{38} \trm{ ergs s}^{-1}$ at the surface. The set of three $T_9$ and $L_{38}$ curves show, from bottom to top, the result if in equation (\ref{eq:entropy}) we take $\epsilon_{\rm nuc} = 0$, $\epsilon_{\rm nuc}=\epsilon_{3\alpha}+\epsilon_{\rm HCNO}$, and $\epsilon_{\rm nuc}=\epsilon_{\rm rp} + \epsilon_{3\alpha}+\epsilon_{\rm HCNO}$. \label{fig:stableburning}}
\end{figure}

If the shock does not trigger unstable $^4$He burning, the H and $^4$He still burn in a stable manner due to the large heat flux from the ashes of $^{12}$C burning below. We account for three sources of stable burning: hot CNO burning $\epsilon_{\rm HCNO} = 5.8\times10^{15} Z_{\rm CNO}\egs$ (we assume the initial mass fraction of CNO nuclei is $Z_{\rm CNO}=0.01$), triple-alpha $^4$He burning $\epsilon_{3\alpha}$, and H burning via the rapid proton capture (rp) process $\epsilon_{\rm rp}$ of \citet{Wallace:81}. The evolution of stable H and $^4$He burning during the first $1000\trm{ s}$ following hydrostatic settling is shown in the top panel of Figure \ref{fig:stableburning}.

To calculate $\epsilon_{\rm rp}$, note that after hydrostatic settling the temperature and density at the base of the H/He layer $y_{\rm He}$ are $T\simeq5\times10^8\trm{ K}$ and $\rho\simeq10^6\trm{ g cm}^{-3}$. The primary breakout reaction from the hot CNO cycle is therefore $^{15}\trm{O}(\alpha, \gamma)^{19}\trm{Ne}$ \citep{Schatz:99}.\footnote{Due to the heat flux from the ashes of $^{12}$C burning, the temperature at $y_{\rm He}$ reaches $T\simeq9\times10^8\trm{ K}$ just $0.01\trm{ s}$ after hydrostatic settling. At this temperature, a second hot CNO cycle emerges that includes $^{18}$Ne \citep{Wiescher:99}, and the primary breakout reaction may instead be $^{18}\trm{Ne}(\alpha, p)^{21}\trm{Na}$ \citep{Schatz:99}.} As the thermal wave from the hot ashes of  $^{12}$C burning diffuses through the H/He layer, the temperature at $y_{\rm He}$ rapidly rises and the $^{19}$Ne captures a proton before it can $\beta$ decay ($t_{\rm 1/2}=17.2\trm{ s}$) and return to the hot CNO cycle. A series of fast proton captures ensue until the nuclear flow reaches the first waiting point of the rp-process, $^{24}$Si ($t_{\rm 1/2}=0.102\trm{ s}$; \citealt{Wiescher:98}). Energy production by rp-processing drops until higher temperatures initiate further proton and $\alpha$ captures. Here we follow Wiescher et al. (1999; see also \citealt{Cooper:06b}) and approximate the energy generation rate of rp-process burning by accounting for just the energy release in burning to $^{24}$Si. Since  $^{15}\trm{O}(\alpha, \gamma)^{19}\trm{Ne}$ is the slowest reaction of the sequence, it governs the total reaction rate of the flow, and
\beq
\label{eq:rpburning}
\epsilon_{\rm rp, 18} = (0.018) X_{\rm He} Z_{\rm CNO} Q_{\rm rp} \rho T_9^{-3/2} e^{-5.85/T_9},
\eeq
in units of $10^{18} \egs$. Here $Q_{\rm rp}=29.96\trm{ MeV}$ is the energy release in burning to $^{24}$Si and the $^{15}\trm{O}(\alpha, \gamma)^{19}\trm{Ne}$ reaction rate is from \citet{Caughlan:88}. This approximation only allows us to place a lower-bound on the contribution of rp-process burning to the light curve; the full calculation is left for future study. Nonetheless, as the bottom panel of Figure \ref{fig:stableburning} shows, it has a noticeable impact on the early time light curve and if the shock does not trigger unstable $^4$He burning, $\epsilon_{\rm rp}/\epsilon_{3\alpha}\approx5 -30$ over the duration of the burst. 
    
\section{Cooling Models and Light Curves}
\label{sec:lightcurves} 

Following the passage of the shock and hydrostatic readjustment, a cooling wave propagates inward from the surface  and the shock-heated layers begin to cool via heat diffusion,
\beq
\label{eq:entropy}
C_p \frac{\partial T}{\partial t} =  \frac{\partial F}{\partial y} + \epsilon_{\rm nuc} - \epsilon_\nu,
\eeq
where the flux $F=\rho K(\partial T/\partial y)$, $\epsilon_{\rm nuc}$ is the nuclear energy generation rate due to any remaining fuel, and $\epsilon_\nu$ is the neutrino energy loss rate. The light curve is set by the final burn temperature profile $T_f(y)$, input as the initial condition of (\ref{eq:entropy}), and the heat flux from any burning $\epsilon_{\rm nuc}$ following hydrostatic settling.

\begin{figure*}
\begin{center}
\epsfig{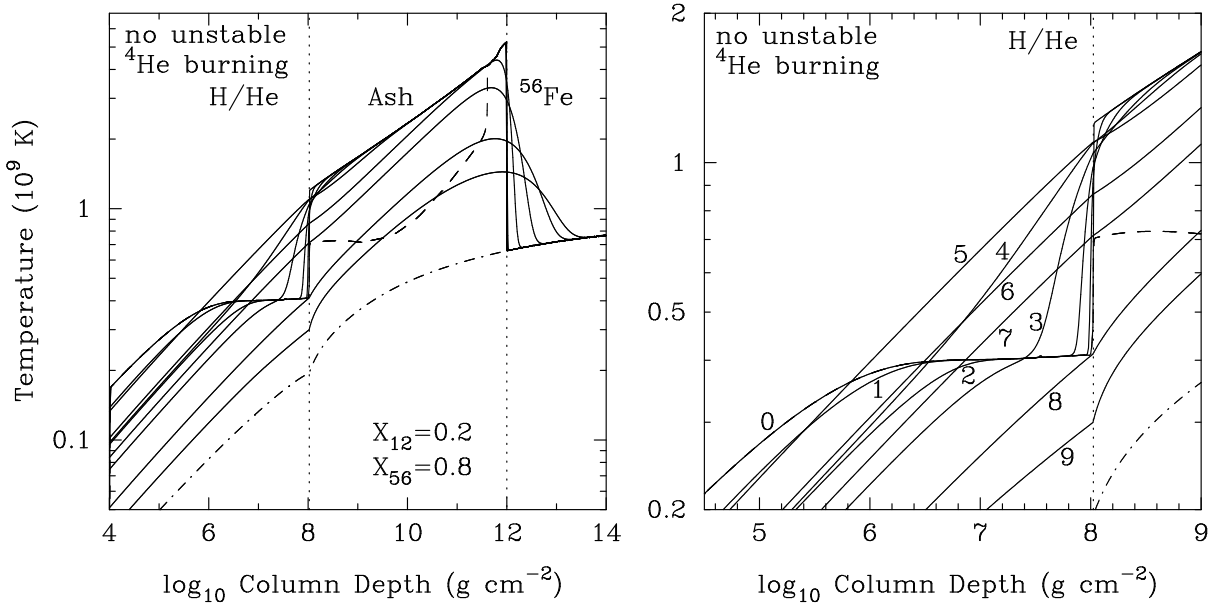}
\end{center}
\caption{Temperature profiles for $y_b=10^{12}\cd$, $\dot{m}=0.1\dot{m}_{\rm Edd}$, and a Type I burst phase $\phi = 0.5$. The mass of freshly accreted H/He is too small for the shock to trigger unstable $^4$He burning (see Figure \ref{fig:trigger}). Shown are profiles at ignition (\emph{dashed-dotted line}),  after the $^{12}$C detonation and shock breakout but before the isobaric deflagration (\emph{dashed line}), and at ten times after hydrostatic settling: $t=0, 0.005, 0.05, 0.3, 1, 10, 1000, 10^4, 6\times10^4, 2\times10^5\trm{ s}$.  The left panel shows the cooling profiles over the entire range in $y$ and the right panel is a zoom in of the H/He layer with the ten profiles labeled in sequential order. The vertical dotted lines show the locations of the H/He---ash and ash---Fe interfaces. \label{fig:TyMol1}}
\end{figure*}

Previous studies assumed $T_f(y)$  was given by the total energy release $\int_{T_i}^{T_f} C_p dT = X_{12}E_{\rm nuc}\simeq (10^{18}\trm{ergs g}^{-1})X_{12}$ assuming an instantaneous and complete isobaric $^{12}$C burn between $10^8\cd < y <y_b$. This gives \citep{Cumming:04b, Cumming:06},
\beq
\label{eq:isobaric}
T_{f, {\rm isobar}} \simeq (5.1\times10^9\trm{ K})\left(\frac{X_{12}}{0.2}\right)^{1/2}\left(\frac{g_{14}y_{12}}{Y_e}\right)^{1/8}.
\eeq
Here we instead use the results of our time-dependent detonation calculation to determine $T_f(y)$.  

The $\epsilon_{\rm nuc}$ term has contributions from two regions of unburned fuel: the $^{12}$C  between $y_{\rm He} < y < y_{\rm det}$ and the H/He at $y < y_{\rm He}$.  As noted in \S~\ref{sec:upward}, due to the sharp temperature gradient at $y_{\rm det}$, the $^{12}$C will likely burn in a convective deflagration that will reach the H/He layer a few $\trm{ms}$ after passage of the shock. For simplicity, we assume an instantaneous isobaric $^{12}$C burn (approximately given by equation [\ref{eq:isobaric}]) in the deflagration region $y_{\rm He} < y < y_{\rm det}$. We assume an energy release in the deflagration of $E_{\rm nuc}=10^{17} \trm{ ergs g}^{-1}$, corresponding to $^{12}$C burning to $^{56}$Fe with $X_{12}\approx0.1$, the fraction of fuel remaining at the onset of the runaway that follows the convective stage.  

To account for H and $^4$He burning, we adopt separate strategies depending on whether the shock triggers unstable $^4$He burning. If the shock triggers unstable $^4$He burning (\S~\ref{sec:unstableHeburn}), the burning layer quickly becomes convective and its thermal evolution during the rise is not described by the heat diffusion equation (\ref{eq:entropy}). Instead, we solve the rise as in Weinberg et al. (2006b; here we account only for $\alpha$-capture reactions), which we then superimpose onto the cooling solution. If the shock does not trigger unstable $^4$He burning, the H and $^4$He still burn in a stable manner (\S~\ref{sec:stableHHeburn}). In this case a convective zone does not form and the thermal evolution is described by equation (\ref{eq:entropy}). We set $\epsilon_{\rm nuc}=\epsilon_{\rm HCNO}+\epsilon_{3\alpha}+ \epsilon_{\rm rp}$, the three sources of stable burning described in \S~\ref{sec:stableHHeburn}, and solve for the rate of change of the composition $dX_i(y)/dt$ simultaneously with the diffusion equation (see Figure \ref{fig:stableburning}).

\begin{figure}
\begin{center}
\epsfig{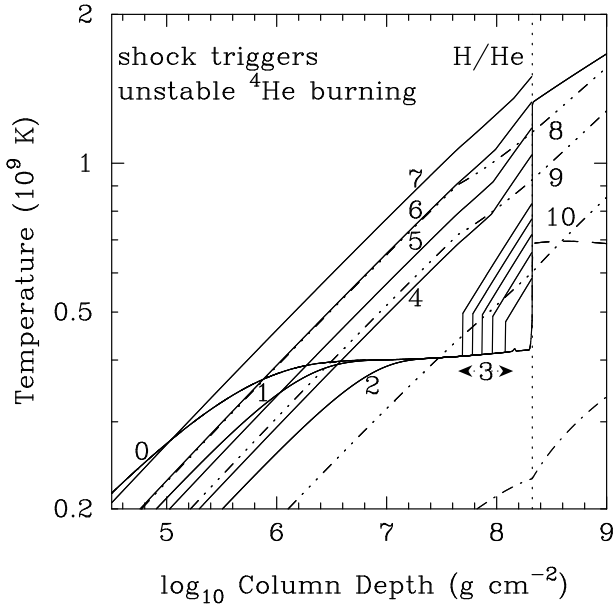}
\end{center}
\caption{Temperature profiles for $y_b=10^{12}\cd$, $\dot{m}=0.1\dot{m}_{\rm Edd}$, and a Type I burst phase $\phi = 1$. Here the shock triggers unstable $^4$He burning, unlike in Figure \ref{fig:TyMol1}. Shown are profiles at several times after hydrostatic settling: Lines $\{0, 1, 2\}$ show the prompt cooling just after shock breakout at times $\{0, 0.01, 0.1\trm{ s}\}$.  Lines $\{3,4,5,6,7\}$ show the H/He layer heating up during $^4$He burning: label 3 shows the growing convective region at $\{0.6,0.9,1.2,1.4,1.7\trm{ s}\}$, and lines $\{4, 5, 6, 7\}$ show the burning as the radiative layers heat up and the flux begins to rise at $\{3.1, 3.9, 4.5, 5.3\trm{ s}\}$. Lines $\{8,9,10\}$ (\emph{dash-triple-dot lines}) show the superburst cooling after $^4$He exhaustion at $\{15, 1000, 3\times10^4\trm{ s}\}$. The dashed-dotted line is the profile at ignition and the dashed line the post-shock, pre-deflagration, profile. \label{fig:TyMol2}}
\end{figure}

We use the method of lines to solve equation (\ref{eq:entropy}), setting $d\ln T/d\ln y = 1/4$ at the outer boundary ($10^4 \cd$) and $dT/dy =0$ (vanishing flux) at the inner boundary ($10^{14}\cd$).  In Figures \ref{fig:TyMol1} and \ref{fig:TyMol2} we show the cooling for $y_{b,12}=1$, $\dot{m} = 0.1 \dot{m}_{\rm Edd}$, and a Type I burst phase $\phi = 0.5$ and $\phi =1$, respectively. In the former case, the shock does not trigger unstable $^4$He burning. The corresponding light curves are shown in Figures \ref{fig:lightcurve2} and \ref{fig:lightcurve1}. We find that in all cases the light curve evolves through three stages: a shock breakout stage ($t\la 1 \trm{ s}$; \S~\ref{sec:breakout}), a precursor stage ($1\trm{ s} \la t \la 20\trm{ s}$; \S~\ref{sec:precursor}), and a superburst stage ($t \ga 20\trm{ s}$; \S~\ref{sec:latetime}). 

\subsection{Shock breakout}
\label{sec:breakout}

For times $\la1 \trm{ s}$ after the detonation, the cooling of the outermost shock-heated layers  ($y\la 10^8\cd$) dominates the emission. It produces a brief ($\approx 0.1\trm{ s}$), bright, flash that precedes the precursor burst peak by $\approx3-10\trm{ s}$ and has an initial luminosity that is super-Eddington. The evolution of the temperature profile within the H/He layer during this cooling epoch is shown in the right panel of Figure \ref{fig:TyMol1} (lines $\{0,1,2\}$). The luminosity of the flash decays as a power-law with time $L\propto t^{-1/2}$, which can be understood as follows (\citealt{Cumming:06} give a similar explanation for the superburst cooling at times $100\trm{ s} \la t \la 1000\trm{ s}$). The cooling wave propagates inward from the surface and reaches a column depth $y_\ast$ at time $t\approx t_{\rm therm}(y_\ast)$, the thermal timescale of the layer. For $y<y_\ast$ the atmosphere has an almost constant heat flux with depth and an approximately constant opacity $\kappa$ so that $L\propto T^4/y=T_f^4(y_\ast)/y_\ast=\trm{ constant}$. For $y>y_\ast$ the temperature profile is nearly unchanged from its initial state $T_f(y)$. Since initially the region $10^6\cd < y < y_{\rm He}$ is nearly isothermal (see \S~\ref{sec:steepening}), $L\propto y_\ast^{-1}$, and since gas pressure dominates $t_{\rm therm} = 3 C_p \kappa y^2 / 4acT^3 \propto y^2$, so that $L \propto t^{-1/2}$.

As the Type I burst phase $\phi \rightarrow 1$, the delay between the onset of the flash and the peak of the precursor burst increases slightly (Figure \ref{fig:lightcurve2}) because the larger $y_{\rm He}$ is, the greater $t_{\rm therm}(y_{\rm He})$ is.  As the $^{12}$C ignition depth $y_b$ increases, the flash becomes more luminous (Figure \ref{fig:lightcurve1}) because the larger $y_b$ is the larger $\Delta p/p$ is at a given depth $y$ (equation [\ref{eq:dpp}]), and thus the hotter the envelope is after hydrostatic settling. 

While the outermost layers $y\la10^7\cd$ cool, the heat flux from the hot ashes of $^{12}$C burning begins to diffuse outward and heat up the layer $10^7 \cd \la y < y_{\rm He}$. A steep temperature gradient develops near $y_{\rm He}$ and a convective zone forms. Since we assume the energy transport is described by heat diffusion (equation [\ref{eq:entropy}]), we underestimate the efficiency of energy transport during this brief ($\la 1\trm{ s}$) convective epoch. The delay between the peak of the shock breakout flash and the peak of the precursor burst may therefore be somewhat shorter than we have found and the valley between the peaks may not be as deep.  
  
Unlike the unstable $^4$He burning precursor, which might be triggered by a shock or an incident $^{12}$C deflagration, the flash is perhaps a unique signature of a $^{12}$C detonation.  A sub-second spike is seen at the onset of the superburst in both cases in which {\it RXTE} caught the rise (4U 1820-30: \citealt{Strohmayer:02}; 4U 1636-54: \citealt{Strohmayer:02b}). In both cases, the peak of the spike precedes the peak of the precursor by $\simeq 2-3\trm{ s}$. Although consistent with the features of a shock breakout, it is possible the spike is instead just the first peak of a photospheric radius expansion burst. However, in the case of 4U 1636-54, the peak flux of the precursor is approximately 60\% fainter than that of the brightest Type I bursts from this system \citep{Kuulkers:04}. This suggests that at least in 4U 1636-54, the precursor is not Eddington-limited and thus not a radius expansion burst. 

\subsection{Precursor burst}
\label{sec:precursor}

\begin{figure}
\begin{center}
\epsfig{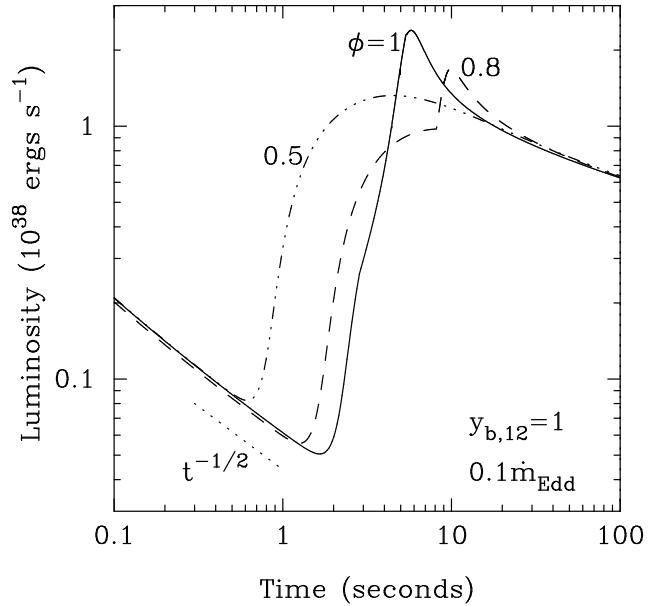}
\end{center}
\caption{Luminosity as a function of time during the early stages of the cooling for an ignition depth $y_{b,12}=1$,  $\dot{m}=0.1\dot{m}_{\rm Edd}$, and a Type I burst phase $\phi = 0.5$ (\emph{dash-triple-dot}), $\phi=0.8$ (\emph{dashed}), and $\phi=1$ (\emph{solid}). For $\phi=0.8$ and $\phi=1$ the shock triggers unstable $^4$He burning. For $\phi=0.5$, H and $^4$He are burned only in a stable manner over $\approx 100\trm{ s}$; approximately $90\%$ of the heat flux at peak is due to the cooling ashes of $^{12}$C burning at depths just below $y_{\rm He}$ (see Figure \ref{fig:stableburning}).\label{fig:lightcurve2}}
\end{figure}

For $1\trm{ s} \la t \la 20\trm{ s}$,  the rise and decay of the unstable $^4$He burning precursor, and the cooling of ashes of $^{12}$C burning just below $y_{\rm He}$, dominate the emission. If unstable $^4$He burning is triggered, it results in a Type I burst that begins to rise $\approx 1\trm{ s}$ after the peak of the shock breakout flash. During the first $\approx1\trm{ s}$ of $^4$He burning, a convective zone develops and moves outward from $y_{\rm He}$ to lower pressure (line 3 in Figure \ref{fig:TyMol2}). The energy generation rate $\epsilon_{3\alpha}$ is so high during the initial stages of burning that the timescale for the convective zone to grow is much shorter than the thermal timescale at the convective-radiative interface. As a result, the outer radiative layers do not get heated during the first second of $^4$He burning and the cooling wave continues to propagate inwards through the outer shock-heated layers.  A steep temperature gradient develops at the convective-radiative interface due to the large compositional contrast between the $^4$He-rich matter that is burning and the outer H-rich material \citep{Weinberg:06}. Eventually, the thermal time at the interface becomes shorter than the convective growth time and the convective zone slowly retreats back to the base $y_{\rm He}$ (lines $\{4,5,6,7\}$ in Figure \ref{fig:TyMol2}). The radiative layers then finally start to heat up and the flux begins to rise, reaching a peak luminosity after several seconds. 

\begin{figure*}
\begin{center}
\epsfig{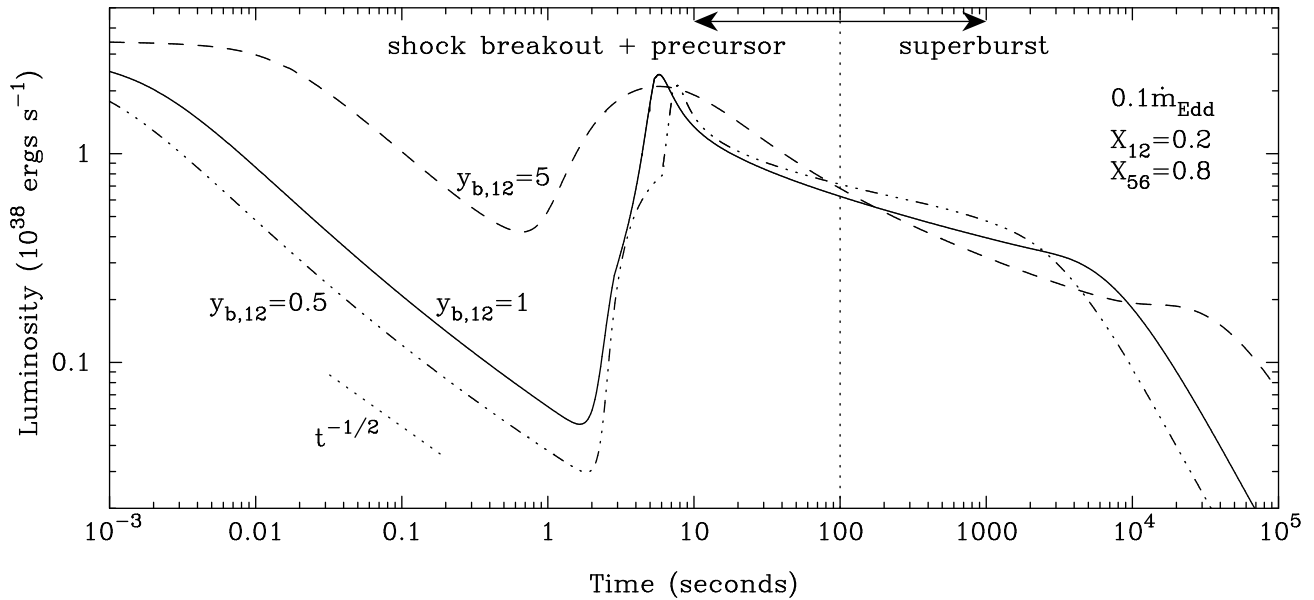}
\end{center}
\caption{Luminosity as a function of time for ignition depths $y_{b,12}=0.5$ (\emph{dash-triple-dot}), 1 (\emph{solid}), and 5 (\emph{dashed}), $\dot{m}=0.1\dot{m}_{\rm Edd}$, and a Type I burst phase $\phi = 1$. In each case, the shock triggers unstable $^4$He burning and the peak of the shock breakout precedes the peak of the precursor burst by a few seconds. The vertical dotted line demarcates where the light curve transitions from the shock breakout and precursor/Type I burst to the superburst. \label{fig:lightcurve1}}
\end{figure*}

The peak luminosity of the Type I burst increases as $\phi \rightarrow 1$ and the amount of $^4$He available to burn increases (Figure \ref{fig:lightcurve2}). After the peak, the H/He layers cool and the thermal wave continues its inward march, gradually penetrating the ashes of $^{12}$C burning (lines $\{8,9,10\}$ in Figure \ref{fig:TyMol2}). 

Importantly, even if unstable $^4$He burning is not triggered, the cooling ashes of $^{12}$C burning just below $y_{\rm He}$ results in a flux rise and decay that is similar in appearance to that of a Type I burst, though with a peak flux that is smaller by a factor of $\approx 2$. This can be seen in the $\phi=0.5$ curve of Figure \ref{fig:lightcurve2}, a case in which the shock fails to trigger unstable $^4$He burning (also compare lines $\{3,4,5\}$ of Figure \ref{fig:TyMol1} with lines $\{3, 4,5,6,7\}$ of Figure \ref{fig:TyMol2}). In this case, the Type I-like rise and decay is powered by the hot ashes of $^{12}$C burning, although \emph{stable} nuclear burning of H and $^4$He also contribute, increasing the peak flux by $\approx10\%-20\%$ (see the bottom panel of Figure \ref{fig:stableburning}; this may be an underestimate since we only account for stable H burning via the rp-process up to $^{24}$Si).  \emph{Thus, even if there is no $^4$He layer, if the $^{12}$C at $y \approx10^8\cd$ burns there will still be a precursor burst, albeit one that is about half as bright as an ordinary Type I burst}.

In three of the five cases in which a precursor was seen, the peak flux of the precursor was smaller than the brightest of the system's ordinary bursts by a factor of $1.5-2$ (the exceptions are the precursor from 4U 1820-30, which looked like an ordinary burst, and 4U 1254-69 which was brighter than an ordinary burst; \citealt{Kuulkers:04, Cumming:06}). It is thus possible that some of the observed precursors are not the result of unstable $^4$He burning. 

\subsection{Late-time light curve}
\label{sec:latetime}
For $t\ga 20\trm{ s}$, the light curve is powered by the continued inward propagation of the cooling wave into the ashes beneath $y_{\rm He}$ (see Figure \ref{fig:TyMol1}, lines $\{6,7,8,9\}$). As \citet{Cumming:04b} showed, the light curve decays as a broken power-law, with the break marking the time when the cooling wave first reaches $y_b$; the deeper $y_b$ is, the later the break (Figure \ref{fig:lightcurve1}). Our final temperature profile $T_f(y)$ in the detonation region  $y_{\rm det} < y < y_b$ is only slightly steeper than the profile in the isobaric deflagration region $y_{\rm He} < y < y_{\rm det}$. Thus, our computed light curves at times $t \ga 100\trm{ s}$ are similar to those of \citet{Cumming:04b}, who assumed an isobaric burn throughout.

Here, as in previous studies, the calculated slope of the pre-break light curve ($t\approx 100-1000 \trm{ s}$) is steeper than the observed slope (see e.g., Figure 6 of \citealt{Cumming:06}).  The emission during this epoch is powered by the cooling of the deflagration-heated region $y_{\rm He} < y < y_{\rm det}$, and in order to better match the observations, $T_f(y)$ in this region must be steeper than we have assumed.  Indeed, since the deflagration is likely to be highly turbulent due to the strong gravity and the hour-long convective churning that precedes the runaway, our assumption of a complete isobaric burn may not be correct. The deflagration may instead leave behind increasingly large pockets of unburned fuel as it accelerates down the density gradient, thereby yielding the steepened temperature gradient implied by the observed light curve.

\section{Summary and Conclusions}
\label{sec:summary}
 
We modeled the hydrodynamic $^{12}$C combustion wave that forms during a superburst as a detonation. We found that the detonation propagates through the deepest layers and drives a shock wave that steepens as it travels upward. Upon reaching the H/He layer, the shock is sufficiently strong that it triggers unstable $^4$He burning if the superburst occurs during the latter half of the Type I burst cycle. We showed that the light curve that results from a shock-triggered $^4$He burn is similar to that of a normal Type I burst. The main difference is that the peak luminosity is somewhat smaller than an ordinary burst's if the superburst occurs early in the second half of the Type I cycle, when there is less $^4$He present.  

A precursor burst has been found in each of the five cases where the onset of the superburst was observed. Given our results, it is unlikely that all are due to shock-triggered unstable $^4$He burning. It is possible that our analysis underestimates the fraction of the cycle over which the shock ignites the $^4$He (see discussion in \S~\ref{sec:trigger}). However, 
because the short thermal time at low densities results in a sharp turnover in the $^4$He ignition curve at $y_{\rm He} < 10^8\cd$, unstable $^4$He ignition is unlikely over much more than half of the cycle, even in a layer of pure $^4$He. A more likely explanation is that some fraction of the observed precursor bursts are due to $^{12}$C burning at depths just below the H/He layer (rp-process H burning may also be stronger than we have found).  We showed that the light curve of a burst in which unstable $^4$He burning is not triggered still has a Type I-like precursor, but with a peak luminosity that is smaller by a factor of $\approx2$. This may explain why the peak luminosity of three of the five observed precursors was about half as bright as the system's brightest ordinary bursts. A larger sample of precursor bursts is needed in order to better test this hypothesis. 
 
Although we modeled the combustion wave as a detonation, the entire $^{12}$C layer may instead burn in a deflagration (see \S~\ref{sec:spontaneous}). How might observations distinguish between the two modes of propagation? The late-time light curve, though well observed, is unlikely to help since at late times the cooling of the ashes of a detonation appears very similar to that of a deflagration (\S~\ref{sec:latetime}). The bright precursor bursts also do not imply a detonation, as a $^{12}$C deflagration may also trigger unstable $^4$He burning when it impacts the H/He layer. Thus, perhaps the most promising signature of a detonation is the shock breakout, a bright, $\approx 0.1\trm{ s}$, flash that precedes the precursor burst by $\approx3-10\trm{ s}$. This may be the origin of the spike seen at the onset of the burst rise in 4U 1820-30 and 4U 1636-54.

The superburst ignition depth is sensitive to the heat release from deep crustal reactions \citep{Brown:04, Cumming:06}. The latter, in turn, depends on the ash composition of previous superbursts \citep{Gupta:06}.  Using a 13 isotope $\alpha$-chain network, we found that the ashes of the detonation consist primarily of $^{28}$Si, $^{32}$S, and $^{36}$Ar. This is potentially interesting since \citet{Gupta:06} find that a crust composed of nuclei with $A = 32-44$, as compared to $A\simeq 50-60$, has a larger heat flux from electron captures. It results in a significantly hotter crust and decreases the superburst ignition depth $y_b$ by a factor of $\approx2$, which helps bring $y_b$ into better agreement with values inferred from observations. Further study is required as our network is small and the results likely depend on the initial composition of the ignition layer (we assumed a $^{12}$C/$^{56}$Fe mixture). For example, superbursts may ignite in a sea of light nuclei if most of the $^{12}$C that fuels the superburst is produced in stable $^4$He burning \citep{Narayan:03,Cooper:05}. Alternatively, superbursts may ignite in a sea of very heavy nuclei  ($A\simeq 100$) made during rp-process H burning \citep{Schatz:99, Schatz:01}. In the latter case, \citet{Schatz:03b} found that the superburst nuclear energy release is dominated by the photodisintegration of the heavy nuclei, which results in a final ash composition that instead consists primarily of iron-group elements.
 
\acknowledgments
We thank F. Timmes for making his reaction network solver available online, C. Fryer and G. Rockefeller for helpful discussions on modeling reactive fluid flow, and E. Brown and R. Cooper for helpful conversations. We also thank the referee, Stan Woosley, for his constructive comments that greatly improved our discussion of the onset of a detonation in this unique environment. This work was supported by the Theoretical Astrophysics Center at Berkeley and the National Science Foundation under grants PHY 05--51164 and AST 02-05956.

\end{document}